\documentclass[12pt]{article}
\usepackage{epsfig}
\usepackage{axodraw}
 
\newcommand{\mrm}[1]{\mathrm{#1}}

\newcommand{\qbar}{\mrm{\overline{q}}}

\newcommand{\pt}{p_{\bot}}
\newcommand{\ptmin}{p_{\bot min}}

\newlength{\abstwidth}
\newlength{\captivewidth}
\begin{document}
 
\sloppy
 
\pagestyle{empty}
 
\begin{flushright}
LU TP 00--01 \\
January 11, 2000
\end{flushright}
 
\vspace{\fill}
 
\begin{center}
{\LARGE\bf A Model for}\\[3mm]
{\LARGE\bf the Colour Form Factor}\\[3mm]
{\LARGE\bf of the Proton}\\[10mm]
{\large  Master of Science Thesis by Johann Dischler}\\ [2mm]
{\large  Thesis advisor: Torbj\"orn Sj\"ostrand}\\ [2mm]
{\it Department of Theoretical Physics,}\\[1mm]
{\it Lund University, Lund, Sweden}
\end{center}
 
\vspace{\fill}
\begin{center}
{\bf Abstract}\\[2ex]
\begin{minipage}{\abstwidth}
The total cross-section $\sigma_{tot}$ and the jet cross-section $\sigma_{jet}$
differ at a proton-proton collision. The latter is divergent if arbitrarily
small transverse momenta are allowed. Even with some fixed lower $\pt$ cutoff,
$\sigma_{jet}$ increases much faster than $\sigma_{tot}$ at high energies.
We have in this paper studied how the
divergence could be tamed by colour screening effects among the partons.

To do this we have built a proton model where we assign momenta, positions and
colour charge to all partons in the proton.

We find that the relative behaviour of the cross-sections can be better 
understood by the inclusion of this effect.
\end{minipage}
\end{center}

\vspace{\fill}
 
\clearpage
\pagestyle{plain}
\setcounter{page}{1}
%
\section{Introduction}
\label{sec-intro}
The Standard Model is the theory that all modern particle physics is
 based on. It tells you which particles that exist and how 
these particles interact with each other. The Model divides the
 particles into two main groups, matter particles and gauge bosons.
As the name indicates, all matter is built up from the matter 
particles. The gauge bosons are the particles that mediate the
interactions ``forces'' between the matter particles and themselves. 
The matter particles are divided in two
different groups, the quarks and the leptons. These, in their turn, 
are placed in groups called families according to: 
%
\begin{equation}
\begin{array}{ccc}
\left( \begin{array}{c} \mrm{u} \\ \mrm{d} \end{array} \right) 
& \left( \begin{array}{c} \mrm{c} \\ \mrm{s} \end{array} \right) 
& \left( \begin{array}{c} \mrm{t} \\ \mrm{b} \end{array} \right) 
\end{array}
\nonumber
\end{equation}
\begin{equation}
\begin{array}{ccc}
\left( \begin{array}{c} \nu_{\mrm{e}} \\ \mrm{e} \end{array} \right) 
& \left( \begin{array}{c} \nu_{\mu} \\ \mu \end{array} \right) 
& \left( \begin{array}{c} \nu_{\tau} \\ \tau \end{array} \right) 
\end{array}
\label{leptons}
\nonumber
\end{equation}
The upper ones are the quarks and they are named up, down, charm, strange,
 top and bottom. The quarks u, c, t all have $(2/3)e$ in electric charge
 and d, s, b have $-(1/3)e$. All the matter that we have around 
 us is made up of only the first family, that is up and down; the rest
of the quarks exists today only inside accelerators and in some very energetic
cosmic objects.

The particles in Eq~(\ref{leptons}) are the leptons. The most famous of 
the leptons is the electron, but there are five more.
The electron $\mrm{e}$, the muon $\mu$ and the tau $\tau$ all have
 electric charge $-e$, while the neutrinos $\nu$ are electrically
 neutral.\\

Then there are the gauge bosons that mediate the different forces
 between the particles. There are four fundamental forces in the universe:
\begin{equation}
\begin{array}{lll}
\underline{\mrm{Interaction}} & \underline{\mrm{Gauge~bosons}} \\
\mrm{gravitation} & \mrm{graviton}\\
\mrm{electromagnetic} & \mrm{\gamma}\\
\mrm{weak} & \mrm{W^{+},~~W^{-},~~Z}^{0}\\
\mrm{strong} & \mrm{g}_{i},~~i=1 \ldots 8.
\end{array}
\nonumber
\end{equation}
Gravitation interacts with all matter, but it is so weak that it
can be neglected almost always in particle physics. Then there is the 
electromagnetic
force which is responsible for holding the atom together. The weak 
force  is e.g. responsible for the $\beta$-decay in the nuclei. In the
 standard model one has succeeded to combine the electromagnetic and 
the weak forces to one---the electro-weak force. And last there is the
 strong force that holds the quarks and nucleons together \cite{ref:hadrons}.
\begin{figure}
\begin{center} 
\begin{picture}(400,100)(0,0)
\Vertex(200,50){1.5}
\ArrowLine(120,50)(200,50)
\ArrowLine(200,50)(260,20)
\Gluon(200,50)(260,80){4}{4}
\put(203,78){$g_{\overline{g}r}$}
\put(203,18){$q_g$}
\put(110,48){$q_r$}
\end{picture}
\end{center}
\caption{A gluon can change the colour of a quark}
\label{fig:gluon-quark}
\end{figure}
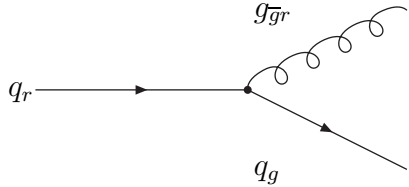 

In this paper we are going to look at the strong force and its 
gauge-bosons, the gluons, so a deeper study of this is in order.
The strong charge is called colour-charge. Instead of $+$ and $-$ 
there is red, blue, green and their anti-colours antired, antiblue and
 antigreen. All matter in the universe is built up of the quarks and the
leptons. The quarks can't exist in isolation, but must be combined in
 colour-singlet states. This can be done in two main ways. You can
  combine a quark (colour) with an anti-quark (its anti-colour), this is
 called a meson, or you can take  one quark of each colour
(e.g. red-green-blue),
this is called a baryon. The gauge bosons of the strong force, the
 gluons, can themselves carry colour charge. This means that the 
emission or absorption of a gluon
 can change the colour of the quarks as seen in 
Fig.~\ref{fig:gluon-quark}. The 
gluons have colour charge either as colour and a different anticolour
(e.g.red-antiblue,$r\overline{b}$) or are in colour 
neutral states (e.g.$\frac{1}{\sqrt{2}}(r\overline{r}-g\overline{g})$).\\

Almost all research in particle physics is done at huge accelerators. 
One is located near Chicago, called the Tevatron, and it is part of a 
 research center called Fermilab. The accelerator ring is
 buried under the ground in a tunnel that has a circumference of about
 6 km. Here  one collides protons (or actually protons and
 antiprotons) at energies as high as $E_{CM}=1.8$ TeV.

It is such collisions that we are going to study in this paper, or a bit
more exactly, the cross-section of proton-proton collisions \cite{ref:cross}.
 The cross-section
is a measure of how probable a reaction is. One can think of it as an
effective radius of the incoming particles. If the two incoming particle's
``radii'' overlap each other we have an interaction. In our case we confront
two different cross-sections, the proton-proton total cross-section 
$\sigma_{tot}$ and the one called $\sigma_{jet}$. In $\sigma_{jet}$ we look 
at the particles inside the proton and their collisions instead. The 
particles inside the proton are usually called partons, and these include both
 quarks and gluons. It is these collisions between the partons that one can
 calculate theoretically. From these calculations one can create a theoretical 
$\sigma_{tot}$ that you can compare with the experimental total cross-section.
The measured cross-section increases much slower with the energy than the
 theoretical. There already exists some explanations to this, such as
 multiple interactions, see Section~\ref{subsec-MultiInt},   but these are 
not enough to explain the gap between the experimental data and the
 theoretical predictions. 

What we have done is to 
build a simple model of a proton with all its particles, partons. We have for
 all these partons calculated the momenta, positions and the 
colour-charges. We have also constructed the model so that the whole 
proton must be colour-neutral. What we are hoping to see is that the partons
 will screen each other and thereby the cross-section will be reduced. If the
energy is high, the number of partons is large and this screening effect 
should be bigger. That is, the increase of the cross-section should be slower,
 which is exactly what is desired. We show that such a screening indeed
appears to be an important physical mechanism.\\
  
This paper is organized as follows: We will in Chapter~2 make some theoretical 
arguments about the evolution of our proton model and discuss the screening 
effect. In Chapter~3 we  will describe our model
 for the proton. Later in Chapter~4 we uses this model to observe how 
a gluon interacts with a proton. And last in Chapter~5 we summarize 
and give some conclusions.   
\clearpage
\section{Theory}
\label{sec-theory}

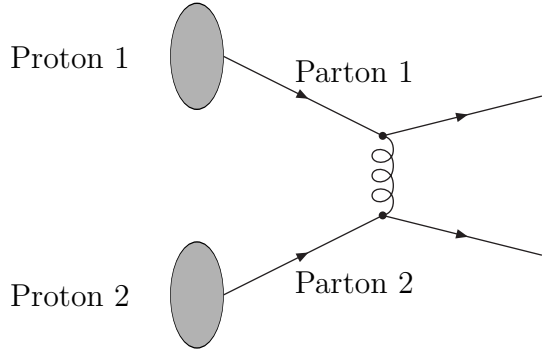
\begin{figure}
\begin{center}
\begin{picture}(400,150)
\Vertex(200,90){1.5}
\Vertex(200,60){1.5}
\Gluon(200,90)(200,60){4}{3}
\ArrowLine(200,90)(260,105)
\ArrowLine(200,60)(260,45)
\ArrowLine(140,30)(200,60)
\ArrowLine(140,120)(200,90)
\GOval(130,120)(20,10)(0){0.7}
\GOval(130,30)(20,10)(0){0.7}
\Text(105,30)[r]{Proton 2}
\Text(105,120)[r]{Proton 1}
\Text(190,115)[]{Parton 1}
\Text(190,35)[]{Parton 2}
\end{picture}
\end{center}
\caption{A proton-proton interaction where two partons scatter against
 each other}
\label{fig:pp-inter}
\end{figure}


If we collide two protons with each other, see Fig.~\ref{fig:pp-inter}
, we have a parton-parton cross-section according to:

\begin{equation}
\mrm{d}\sigma_{jet} = \int \mrm{d}x_{1}f(x_{1},Q^2)\int \mrm{d}x_{2}f(x_{2},
Q^2)\int \frac{\mrm{d}\hat{\sigma}}{\mrm{d}\pt^2}\mrm{d}\pt^2.
\label{cross-sectq}
\end{equation}

The first integral factor in Eq~(\ref{cross-sectq}) expresses the flux of 
partons from proton~1. The second term is then the flux of partons from
 proton~2. The last term is a measure of the actual probability that two
 partons will interact with each other. 

The parton function $f(x,Q^2)$ expresses the probability to find a parton
with momentum fraction $x$ of the incident proton momentum, while Q sets
the virtuality scale at which the proton is probed.
Q is often in hadron-hadron collisions represented by $\pt$ since these
 variables scale approximately the same, that is
\begin{equation}
\mrm{d}\sigma_{jet} = \int \mrm{d}x_{1}f(x_{1},\pt^2)\int \mrm{d}x_{2}
f(x_{2},\pt^2)
\int \frac{\mrm{d}\hat{\sigma}}{\mrm{d}\pt^2}\mrm{d}\pt^2.
\label{cross-sectp}
\end{equation}
The parton functions $f(x_i,\pt^2)$ in Eq~(\ref{cross-sectp}) increase 
very fast at small $x$-values (see Section~\ref{subsec-eveq} for further
 details), that is small momentum for the partons. They also
 increase for bigger virtuality, Q. This will eventually lead to a 
 parton-parton cross-section that is bigger than the whole proton-proton 
cross-section.
\clearpage

\subsection{Evolution Equations}
\label{subsec-eveq}

When you are studying how a hadron, e.g a proton, is built up you have to do
this by sending a particle, a probe, at the hadron. This probe then scatters
against a parton inside the hadron and by this you can draw conclusions
about how the hadron is made up. If the probe has a large $Q$ (is very 
virtual) it can resolve more details and we see more partons.

That there are details to resolve is a consequence of the quantum mechanical
fluctuations. Virtual fluctuations $a \rightarrow b c \rightarrow a$ appear and
disappear continuously, with lifetimes related to the virtuality scale according
to the Heisenberg uncertainty principles. The fluctuations can be nested, 
thereby giving rise to whole cascades of fluctuations.

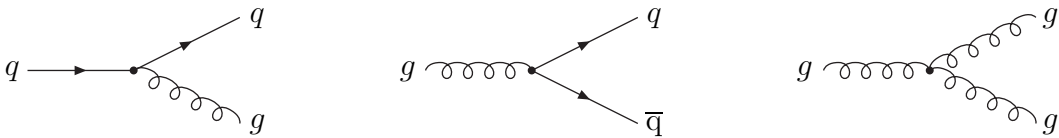
\begin{figure}[h]
\begin{center} 
\begin{picture}(400,100)(0,0)
\Vertex(200,50){1.5}
\Vertex(50,50){1.5}
\Vertex(350,50){1.5}
\ArrowLine(10,50)(50,50)
\ArrowLine(50,50)(90,70)
\Gluon(50,50)(90,30){3}{4}

\Gluon(160,50)(200,50){3}{4}
\ArrowLine(200,50)(240,70)
\ArrowLine(200,50)(240,30)

\Gluon(310,50)(350,50){3}{4}
\Gluon(350,50)(390,70){3}{4}
\Gluon(350,50)(390,30){3}{4}

\Text(7,50)[r]{$q$}
\Text(97,70)[]{$q$}
\Text(97,30)[]{$g$}
\Text(157,50)[r]{$g$}
\Text(247,70)[]{$q$}
\Text(247,30)[]{$\qbar$}
\Text(307,50)[r]{$g$}
\Text(397,70)[]{$g$}
\Text(397,30)[]{$g$}
\end{picture}
\end{center}  
\caption{The 3 main vertices in the strong interaction}
\label{fig:3vertices}
\end{figure}

There are three different processes that can happen in the parton cascade.
These are the three vertices of the strong interaction of the first order.
 A quark can split into a quark and a gluon. And a gluon can either 
split into a quark-antiquark pair or into two gluons, see 
Fig.~\ref{fig:3vertices}. These processes are of course reversible, so e.g 
a quark can absorb a gluon. 

The daughters share the momentum of the mother between themselves 
according to some probability distributions. It is e.g. a large probability 
that the gluon emitted by the quark has very small momentum. These probability
 distributions are described by the Altarelli-Parisi splitting kernels 
\cite{ref:Altarelli}

\begin{eqnarray}
P_{\mrm{q}\rightarrow\mrm{qg}}(z) &=& \frac{4}{3} \, \frac{1+z^2}{1-z},
\label{P1}
\\
P_{\mrm{g}\rightarrow\mrm{gg}}(z) &=& 3 \, \frac{(1-z(1-z))^2}{z(1-z)},
\label{P2}
\\
P_{\mrm{g}\rightarrow\mrm{q}\qbar}(z) &=& n_{\mrm{f}} \, \frac{1}{2} \, 
(z^2+(1-z)^2),
\label{P3}
\end{eqnarray}
where $n_{\mrm{f}}$ is the number of allowed flavors, usually 3-5. The variable
$z$ is the fraction of the mother momentum that the first daughter
takes (i.e. the quark in Eq~(\ref{P1})). The other daughter then gets 
$(1-z)$ of the momentum. Both in Eq~(\ref{P1}) and Eq~(\ref{P2}) there
are singularities, in Eq~(\ref{P1}) at $z=1$ and in Eq~(\ref{P2}) both at
$z=1$ and $z=0$. The physical explanation is that there, at any given time,
is a cloud of gluons with very low momenta around a quark. In reality 
even if there are are infinitely many gluons so are their total momentum
finite. One usually introduce a cutoff, $z_{min}$, and branchings that includes
gluons below this $z_{min}$  (or over $(1-z_{min})$) are then not allowed.

The probability that we will have a branching 
$\mrm{a}\rightarrow\mrm{b}\mrm{c}$ at all is given by the evolution equation,
the DGLAP equation \cite{ref:Altarelli}:

\begin{equation}
\frac{\mrm{d}P_{\mrm{a}\rightarrow\mrm{b}\mrm{c}}}{\mrm{d}(\mrm{ln} Q^2)}=\int 
\mrm{d}z\frac{\alpha_s(Q^2)}{2\pi}P_{\mrm{a}\rightarrow\mrm{b}\mrm{c}}(z)
\label{eq:exc1}
\end{equation}
or, if one introduces the evolution parameter $t=\ln\frac{Q^2}{\Lambda^2}$ 
\begin{equation}
\frac{\mrm{d}P_{\mrm{a}\rightarrow\mrm{b}\mrm{c}}}{\mrm{d}t}=\int 
\mrm{d}z\frac{\alpha_s(t)}{2\pi}P_{\mrm{a}\rightarrow\mrm{b}\mrm{c}}(z).
\label{eq:exc2}
\end{equation}
That is, for a small increase in $t$, d$t$, there is a probability d$P$ that a
branching  $\mrm{a}\rightarrow\mrm{b}\mrm{c}$ will take place. So the 
evolution equation evolves in the virtuality, $Q$. This does not mean that 
the individual partons evolve in Q, but it is simply an easy way to pick the
 right value for the branchings at which the partons are resolved.\\

Another way of expressing this is an inclusive picture; you could calculate
the probability of finding a parton at a certain $x$ value. This becomes 
more like a mean-value of the distribution, but it is a picture that is 
often used since it is these distributions that you can measure at experiments,
 and not how a single parton develops. We have, for the inclusive picture, 
the following expression:
\begin{equation}
\frac{\mrm{d}f_i(x)}{\mrm{d}(\ln Q^2)}=\int_x^1 \mrm{d}yf_j(y)\int_0^1 
\mrm{d}z\frac{\alpha_s}{2\pi}P_{\mrm{j}\rightarrow\mrm{i}\mrm{k}}(z)\,
\delta(yz-x).
\label{eq:ink1}
\end{equation}
The first integral in Eq~(\ref{eq:ink1}) simply says that the mother, $j$, 
has to have at least $x$ momentum. The other integral one recognizes
from our earlier expression in Eq~(\ref{eq:exc1}), and the delta function
ensures that the daughter, $i$, obtains $x$ in momentum. If we use the 
delta function to eliminate one of the integrals we get
\begin{equation}
\frac{\mrm{d}f_i(x)}{\mrm{d}(\ln Q^2)}=\int_x^1 \frac{\mrm{d}y}{y}f_j(y)
\frac{\alpha_s}{2\pi}P_{\mrm{j}\rightarrow\mrm{i}\mrm{k}}
\left(\frac{x}{y}\right).
\label{eq:ink2}
\end{equation}
So this expresses number of daughters that will get $x$ momentum from a
mother of momentum higher than $x$. But these partons with $x$ momentum
can also decay, so we have also an out going flow from this point that 
has to be subtracted from the expression. 

Once you have a parton distribution at some $Q$-scale you can by these
evolution equations iterate towards higher $Q$ and get the whole distribution
above this $Q$. This has been
done by many groups \cite{ref:MRST,ref:CTEQ} for different starting 
values. One feature that is common for all groups is that the distributions 
are very high at small $x$-values.

\subsection{Multiple Interaction}
\label{subsec-MultiInt}

If two protons collide with each other it can happen that more than one 
pair of partons collides inside the protons. This we call multiple
interaction \cite{ref:multi}. There are some experimental data that indicates
this phenomena \cite{ref:CDF,ref:UA2}.

\begin{figure}[h]
\begin{center}
\rotatebox{270}{\mbox{\epsfig{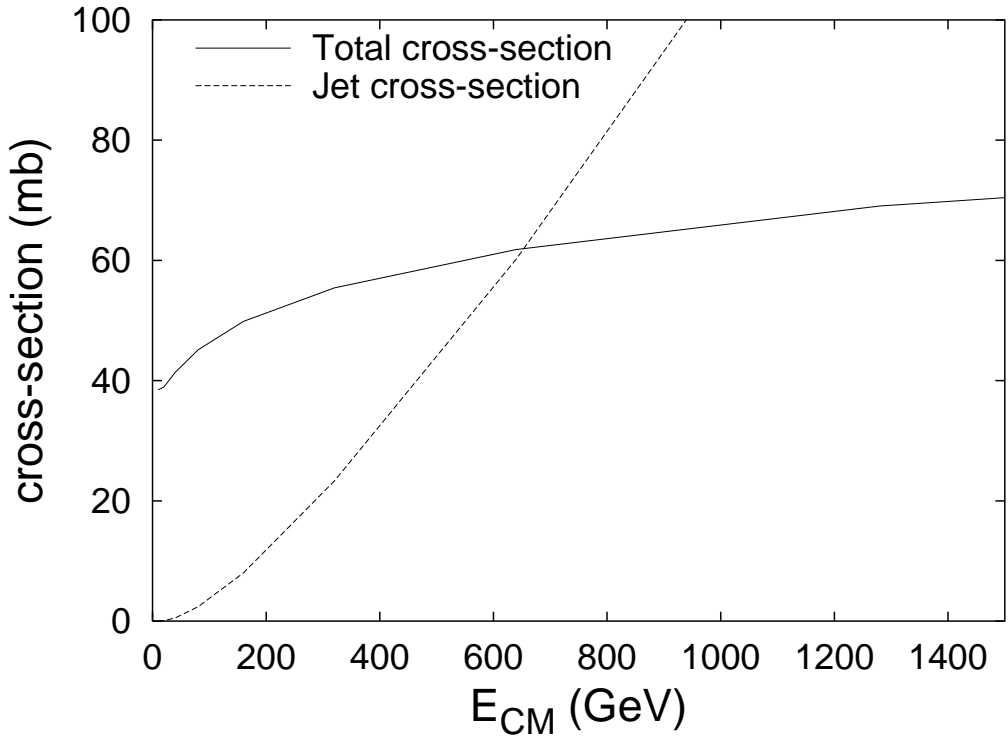}}}
\end{center}
\caption{The total cross-section $\sigma_{tot}$ and the parton-parton cross 
section $\sigma_{jet}$. $\ptmin$ is here set equal to 2 GeV }
\label{fig:traffyta} 
\end{figure}

The total cross-section, $\sigma_{tot}$, is of the order $40-70$ mb see
Fig.~\ref{fig:traffyta}. It is often parameterized according to 
\cite{ref:sigma}:
\begin{equation}
\sigma_{tot}=As^{-\eta}+Bs^{\epsilon},
\end{equation}
where $s$ is the $E_{CM}$ squared, $\eta\sim0.5$ and $\epsilon\sim0.08$. 

Let us, on the other hand, look at the parton-parton cross-section, and make
some rough estimates such as: $f_i(x,t)\sim x^{(-1-\epsilon')}$,
and introduce an $x_{min}$ to avoid to get singularities at $x=0$
and finally a $\ptmin$. $\ptmin$ is introduced as an integration limit that
you after calculating the integral should put to zero. If you do all this you 
will get something like:
\begin{eqnarray}
\sigma_{jet} & = & \int \mrm{d}x_{1}f(x_{1},Q^2)\int \mrm{d}x_{2}f(x_{2},
Q^2)\int \frac{\mrm{d}\hat{\sigma}}{\mrm{d}\pt^2}\mrm{d}\pt^2 \nonumber \\ 
 & \sim &
\left(\frac{1}{x_{min}^{\epsilon'}}\right)^2\frac{1}{\ptmin^2} \sim
\frac{s^{\epsilon'}}{\ptmin^2}.
\end{eqnarray}
As is clearly seen, one must have $\ptmin \neq 0$. This contradiction can, we
hope, be explained by the screening effect, see the next section. 
If you do these calculations a bit more carefully, or uses a simulation 
program such as PYTHIA 6.1 \cite{ref:PYTHIA} you get curve according to 
Fig.~\ref{fig:traffyta}.
One can increase $\ptmin$ but this will only move the problem to another
region.
%
%

The huge difference between the two curves can, in part, be explained by 
multiple interaction since we have:
\begin{eqnarray}
\sigma_{tot} & = & \sigma_{0jetpair} +\sigma_{1jetpair}+\sigma_{2jetpairs}+
\cdots,
\\
\sigma_{jet} & = & \sigma_{1jetpair}+2\sigma_{2jetpairs}+3\sigma_{3jetpairs}+
\cdots,
\end{eqnarray}
where we have to multiply by the number of jets in $\sigma_{jet}$ because
$\sigma_{ijetpair}$ tells about the probability to have $i$ jet pairs. If we 
have $i$ jet pairs we always has $i$ such terms.This leads
to that at an event with many jets (high energy) the $\sigma_{jet}$ will
be higher than $\sigma_{tot}$.

This seems to explain the curves well, but unfortunately it's not enough. 
We believe that the necessary last correction can be explained by the 
screening effect,
that is that the quarks will not interact as if they are free particles 
which has been assumed in the perturbation calculations of the $\sigma_{jet}$
cross-section. 
\subsection{Screening effect}
\label{subsec-screening}
All hadrons are in colour singlet states. This means that if you can't
resolve a hadron it will not interact strongly. That is if you e.g. send a
gluon with a very long wavelength at a proton, the gluon will not be able
to see the partons inside the proton, but only the proton as a single 
object. Since this object is in a colour singlet state the gluon will have 
problem to interact with it. On the other hand, if the gluon is very 
energetic and has a short wavelength, it will resolve much more partons and
the screening effect will become much smaller.

This can be studied by looking at the ratio $A$ between the incoherent and the
coherent sum of colour charges in the cascade.
\begin{equation}  
A=\frac{|\sum_{k=1}^n q_ke^{ipx_k}|^2}{\sum_{k=1}^n|q_k|^2}.
\end{equation}

\begin{figure}[h]
\begin{center} 
\begin{picture}(300,100)(0,0)
\Vertex(100,50){2}
\Vertex(150,50){2}
\Vertex(175,50){2}
\Curve{(50,40)(79,55)(90,60)(117,69)(143,77)(160,81)(195,85)
(241,82)(293,69)}
\Curve{(50,40)(61,59)(73,77)(85,88)(90,90)(101,87)(113,76)(131,49)
(148,23)(160,14)(166,14)(171,15)(177,19)(188,32)(206,60)(218,77)}
\Text(50,50)[rb]{gluons}
\Text(105,47)[rt]{$k=1$} 
\Text(155,47)[rt]{$2$} 
\Text(180,47)[rt]{$3$} 
\Text(210,47)[l]{ $p$ large} 
\Text(240,95)[l]{$p$ small} 
\end{picture}
\end{center}
\caption{The screening effect; A gluon interacts with the partons in a proton 
and depending on the wavelength it interacts differently.}
\label{fig:Amp}
\end{figure}
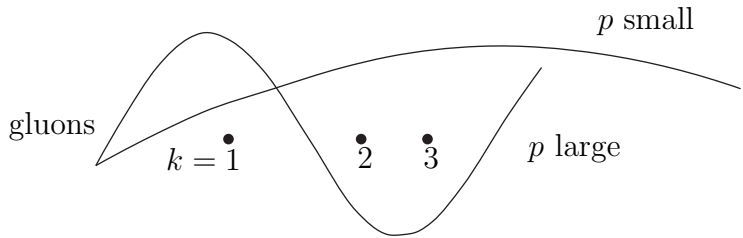

Here is $n$ the number of partons in the proton, $q_k$ is the colour
charge of the $k$:th parton and $x_k$ is the position of the $k$:th parton.
The gluon has a momentum of $p$. If the gluon has a low $p$ the wavelength
is so long that it won't matter where in the proton the partons are, see
Fig~\ref{fig:Amp}, and all terms will cancel. Or you could see it like this:
if $p\rightarrow 0$ then $A\rightarrow 0$ (since $\sum q_k=0$) and if 
$p\rightarrow \infty$  then $A\rightarrow 1$

\clearpage

\section{The Proton-Model}
\label{sec-model}
We start, in our model of the proton, at a very low virtuality scale $Q$.
At this low $Q$ we assume that we have a finite number of partons, or more
specificly, that we have 3 quarks and 2 gluons \cite{ref:grv}. These 5 partons 
are then
evolved to higher $Q$ by Eq~(\ref{eq:exc2}) to some $Q_{max}$. As we
do this we continuously assign each new parton with momentum, position and 
colour charge according to some given probability. We have decided the
transversal position of the partons while in the momentum space the 
longitudinal momentum is selected. We do this since it allows us to build a 
model and still be consistent with the 
Heisenberg Uncertainty Principle. To have a colour-neutral
proton we, to begin with, have a colour-neutral start configuration, but then
as we evolve we keep all partons. To avoid getting too many partons we had to 
introduce a $x_{min}$ and branchings in which a parton had a momentum
under this were not allowed. This rather abrupt cutoff did not change the 
distribution significantly, as we shall see.

\subsection{Start configuration in momentum space}
\label{sec-begmom}
We choose to start with 3 quarks and 2 gluons. In this choice  we were
guided by other works done earlier. In the original in 1976
\cite{ref:urspgrv} one started with only 
the three valence quarks, and this was not enough to fit the data with
experiments. The GRV group \cite{ref:grv,ref:grv89} later made the assumption 
that you had to introduce
some gluons and seaquarks already from the beginning, though only very 
few, and this seems to be sensible since they got a good fit to the 
experiments . We cannot use the 
GRV distribution itself, since they use a
inclusive picture, while we need an exclusive picture to be able to assign
momenta, colour charge etc. to all partons. \\

In our model, each parton was  given a fraction $x$ of the total momentum 
from a distribution that was fitted to give values similar to what the GRV
group had. What we are trying to get is a qualitative picture of the parton 
distribution, not to compete with GRV for the best configuration. 
  We made the following ansatz:
\begin{equation}
f(x)=Nx^{\alpha}(1-x)^{\beta},
\end{equation}
where $N$ is a normalization factor to get the right number of partons and
 $\alpha$ and $\beta$ are the free parameters that we used to adjust the distribution.

To get the proper normalization $\sum_{i=1}^5 x_i=1$ we made the following
operation:
\begin{equation}
(x_{i})_{\mrm{norm}}=\frac{x_i}{\sum_{j=1}^5 x_j}.
\end{equation}
This means that the original distribution is pushed towards the middle.
As it were we had to choose $\alpha<0$. Some trial and error finally gave
that:
\begin{eqnarray}
\alpha & = & -0.4 \nonumber \\
\beta & = & 1.2   \nonumber
\end{eqnarray}
%
\begin{figure}
\begin{center}
\rotatebox{270}{\mbox{\epsfig{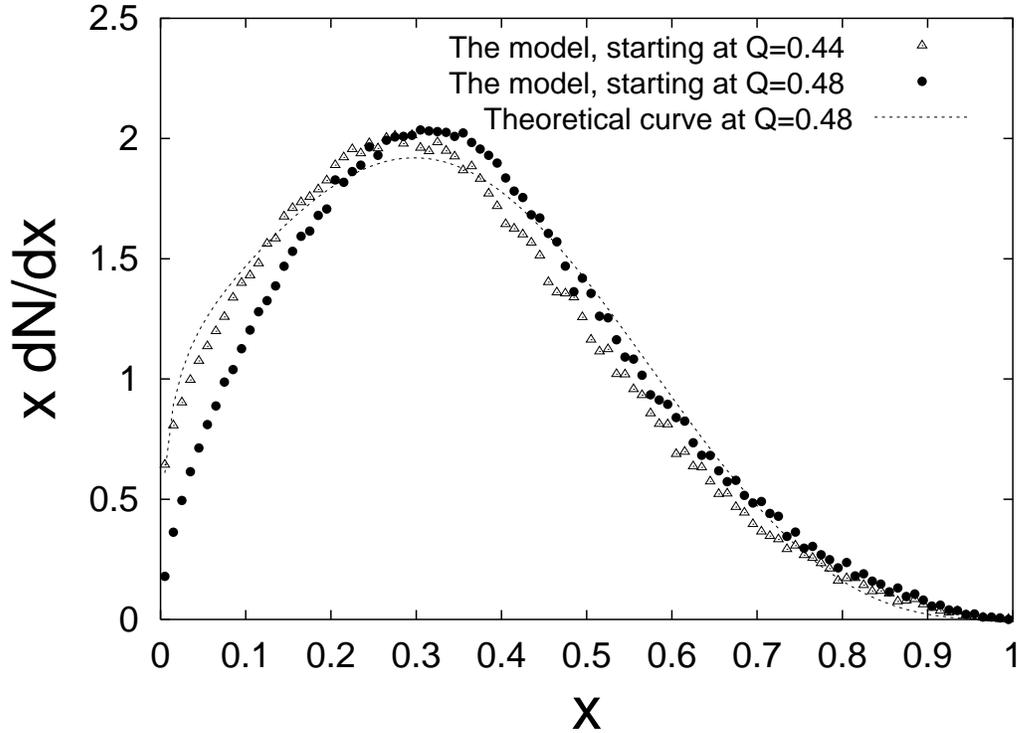}}}
\end{center}
\caption{Different start configurations for the momentum. Theoretical curve
from \cite{ref:grv}}
\label{fig:begmom}
\end{figure}
%
Though even this did not give a very good distribution. We had too much 
partons at high momentum and too few at low, see Fig.~\ref{fig:begmom}. This 
we solved by starting our evolution at $Q=0.44$ instead of $Q=0.48$ as 
they had done.
 That means that when we were at $Q=0.48$ some of the partons with high
momentum had split to two other partons with lower momentum and we got 
a good fit, see Fig.~\ref{fig:begmom}.

\subsection{Start configuration in coordinate space}
\label{sec-begpos}

\begin{figure}[h]
\begin{center} 
\begin{picture}(400,120)(0,0)
\LongArrow(0,50)(100,50)
\LongArrow(50,0)(50,100)
\Oval(50,50)(40,40)(0)
\Line(50,50)(65,85)
\Vertex(65,85){2}
\LongArrow(65,85)(65,51)
\Line(120,20)(120,80)
\Line(140,20)(140,80)
\Line(160,20)(160,80)
\Line(180,20)(180,80)
\Text(103,50)[l]{x}
\Text(50,103)[b]{y}
\Text(110,95)[l]{Incoming plane wave}
\Text(55,75)[]{r}
\end{picture}
\end{center}
\caption{The projection of the position}
\label{fig:projpos}
\end{figure}
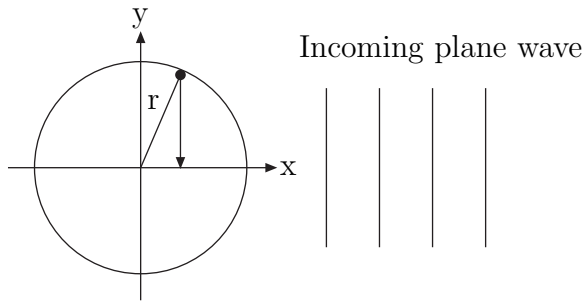

Our model is 2-dimensional. This is not a strong limitation, though, 
since the protons have a very high speed and thereby are strongly 
Lorentz-contracted. That is, they can with good approximation be taken to be
 2-dimensional objects. And since the only thing that we are going to do is 
to send plane waves in the transverse direction of the beam at the proton, 
see Fig.~\ref{fig:projpos}, we can project the points onto the x-axis. So we
put $x=r \cos \theta$ and chose $\theta$ at random for each new branching. 

We have introduced a characteristic radius, $r_0$, of
 the proton at $r_{0}=0.7$ fm. The 5 original partons are then randomly
chosen according to some distribution. We tried some different distributions:
\begin{itemize}
  \item Gauss :\, $f(r) \mrm{d}r \propto 
 \frac{1}{\sqrt{2\pi}r_0} e^ {\frac{-r^2}{2r_0^2}}r \mrm{d}r$  
  \item Exponential I:\,  $f(r)\mrm{d}r  \propto e^{-\frac{r}{r_0}}r\mrm{d}r $
  \item Exponential II :\, $f(r)\mrm{d}r  \propto \frac{e^{-\frac{r}{r_0}}}{r}
r \mrm{d}r $ 
\end{itemize}
Here is $r_0$ the characteristic radius. The difference
 in the final results
of the parton shower did not depend much on which of these distributions 
we choose. We decided to take the Gaussian distribution. 

\subsection{The evolution of the momentum}
\label{sec-evmom}
We have chosen to work with an exclusive picture, that is we follow
the development of each parton in the proton. This picture is easy to
work with since you can iterate from a start configuration of partons.
You know the probability for a parton to split up at a little step, 
$\mrm{d}t$, of the evolution parameter. So by using random numbers to 
simulate the probability of the splitting you can reproduce the actual 
process. In the exclusive picture we had:

\begin{equation}
\frac{\mrm{d}P_{\mrm{a}\rightarrow\mrm{b}\mrm{c}}}{\mrm{d}t}=
\int_{z_{min}}^{1-z_{min}} 
\mrm{d}z\frac{\alpha_s(t)}{2\pi}P_{\mrm{a}\rightarrow\mrm{b}\mrm{c}}(z),
\label{eq:Mexc1}
\end{equation}
with 
\begin{equation}
\alpha_s = \frac{12\pi}{(33-2n_f)\ln(Q^2/\Lambda^2)}=\frac{12\pi}{25t},
\end{equation}
where $n_f=4$ and $t=ln(Q^2/\Lambda^2)$. Since we have that $z_{min}=x_{min}/x$
we can't have any partons under $x_{min}$. For the quark channel 
$q\rightarrow qg$ this gives, with help of the Altarelli-Parisi splitting
 kernels to decide $P_{\mrm{q}\rightarrow\mrm{q}\mrm{g}}$, that the evolution
equation looks as:
\begin{equation}
\frac{\mrm{d}P_{\mrm{q}\rightarrow\mrm{q}\mrm{g}}}{\mrm{d}t}=
\frac{6}{25t} \int_{z_{min}}^{1-z_{min}}  \mrm{d}z \frac{4}{3} 
\frac{1+z^2}{1-z}\equiv \frac{C}{t}.
\label{eq:Mexc2}
\end{equation}
You can, by simply putting in the different splitting kernels, get the 
evolution equations for $g\rightarrow q\overline{q}$ and $g\rightarrow gg$. 
For a gluon you have  to decide which of the channels it will take: either the 
$gg$ or the $q\overline{q}$. This is randomly chosen according to the 
different probabilities of the channels.

The number of partons, $N$, at a given $Q$ is then given by
\begin{equation}
\frac{\mrm{d}N}{\mrm{d}t}=-\frac{C}{t}N(t)
\label{eq:decay}
\end{equation}
One can draw a parallel here to the exponential decay of a radioactive 
nucleon. It has a certain probability, $P$, to decay, but if you look at 
a large number of nucleons the probability that you will have a decay in
a little time interval decreases as the time elapses since you have fewer
nucleons left which can decay. By solving Eq~(\ref{eq:decay})
you get how many partons, $N$, you have at a given Q-scale.

In this way you evolve your parton distribution until you reach $Q=Q_{max}$.
 We have also introduced a $x_{min}$ to keep the number of partons in our
model from exploding. All branchings in which  any of the daughters
has a momentum less than $x_{min}$ are forbidden.

\begin{figure}[h]
\begin{center}
\rotatebox{270}{\mbox{\epsfig{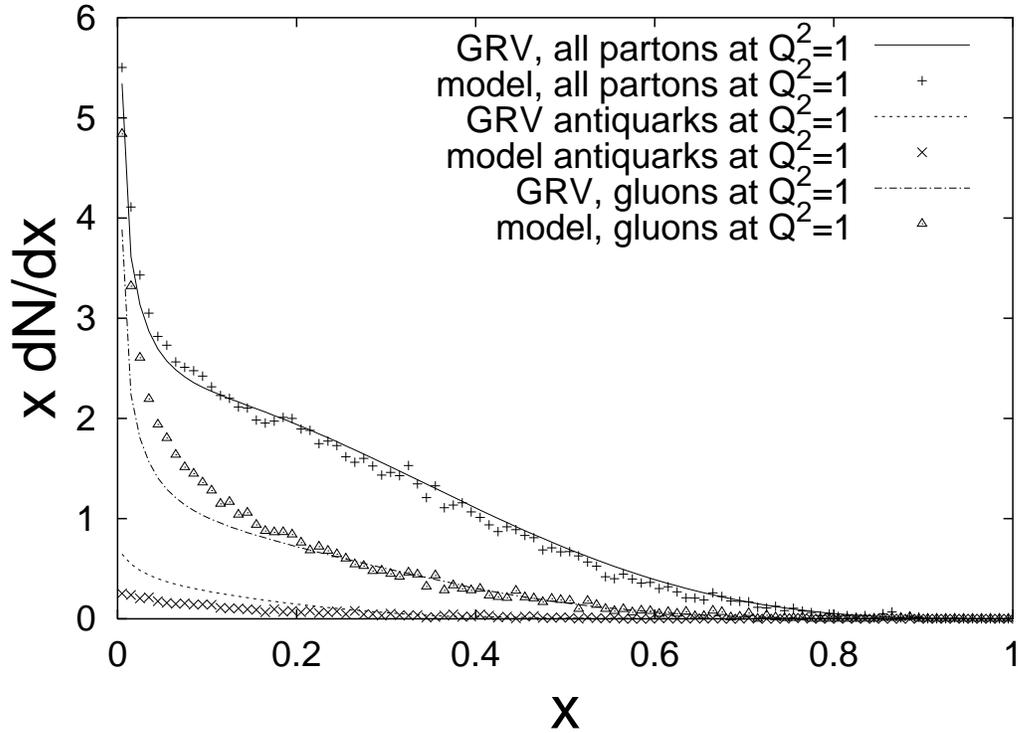}}}
\end{center}
\caption{The evolution of the momentum at $Q^2=1$. At $Q^2=1$
 $\qbar$ is too low in our model. The quark distribution can be attained by
subtracting the gluon and the antiquark distributions from the total.}
\label{fig:evmom1}
\end{figure}

\begin{figure}[p]
\begin{center}
\rotatebox{270}{\mbox{\epsfig{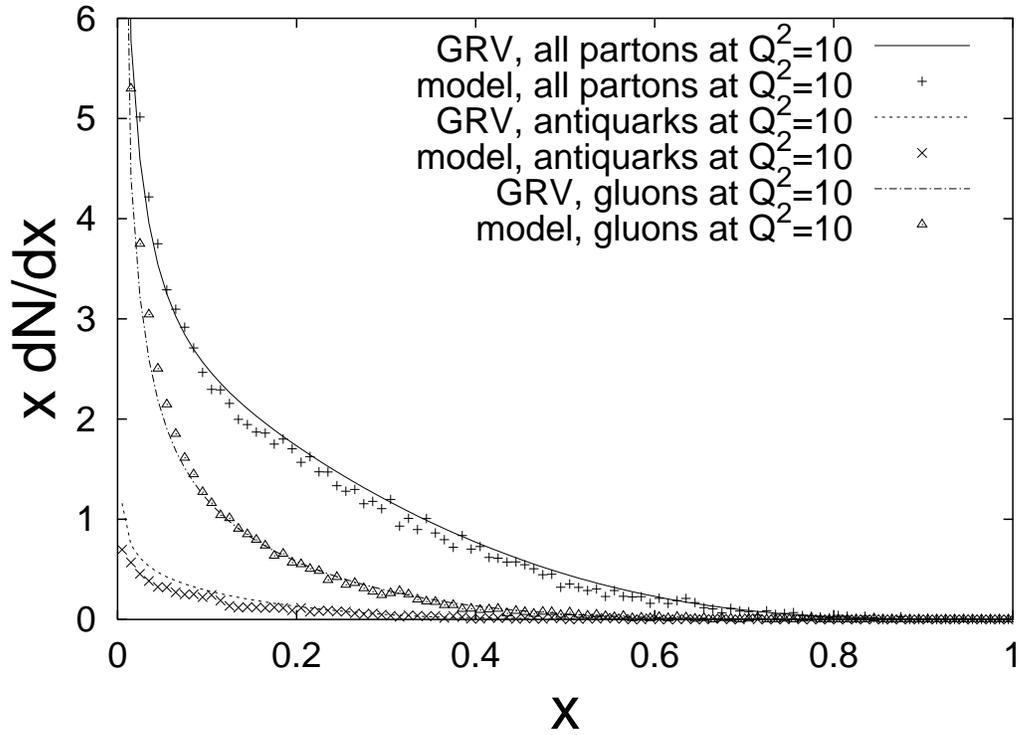}}}
\end{center}
\caption{The evolution of the momentum at $Q^2=10$. The $\qbar$ curve of our
model has approached the theoretical curve.}
\label{fig:evmom10}
\end{figure}

\begin{figure}[p]
\begin{center}
\rotatebox{270}{\mbox{\epsfig{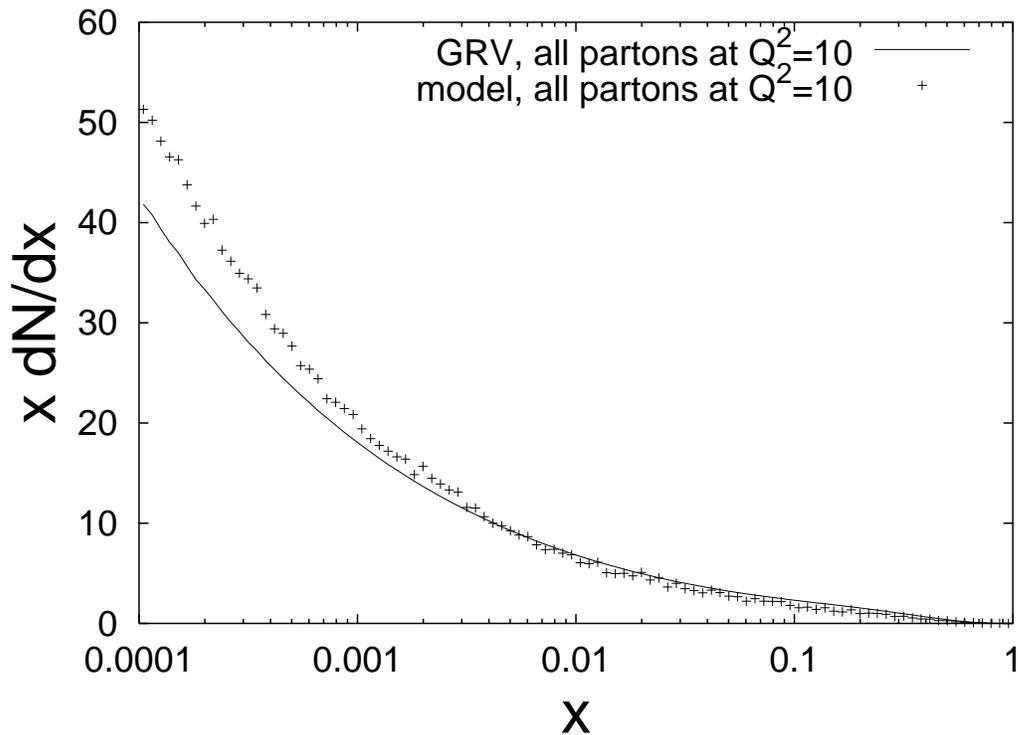}}}
\end{center}
\caption{The evolution of the momentum at $Q^2=10$ in log scale. 
$x_{min}=10^{-4}$}
\label{fig:evmomlog}
\end{figure}

We have compared our model with values from the GRV group \cite{ref:grv}, see 
Fig.~\ref{fig:evmom1} and Fig.~\ref{fig:evmom10}. We have an excellent 
agreement if we include all partons, as seen in the upper curves
. The antiquarks though are somewhat suppressed 
in our model, especially at low $Q$. This is not so strange, since we in 
our model assume that we don't have any antiquarks in our start configuration
 and GRV have a mixture of them. At higher $Q$ these differences shrink and 
the two curves approaches each other, see Fig.~\ref{fig:evmom10} .
We have a fairly good agreement at low $x$-values also, see 
Fig.~\ref{fig:evmomlog}, and this is a bit surprising since we have introduced
a $x_{min}$ cut.
%
%
%
%
%
%
\clearpage

\subsection{The evolution of the position}
\label{subsec-evpos}
 The distance a parton will reach from its origin of creation depends on a 
number of things. In our model we assign each new pair of partons with a 
distance, $x$, according to:
\begin{equation}
x=x_0 \pm \frac{\mrm{R}}{Q}\cos{\theta}\sqrt{1-z_i}
\label{eq:evpos}
\end{equation}
Here is $x_0$ the position of the vertex where the two partons were created 
and R is a random number between 0 and 1.

The first factor $\frac{1}{Q}$ comes from the Heisenberg Uncertainty 
 Principle. A parton with high virtuality, $Q$, lives shorter than a parton
that is almost real. The random number, $R$, is 
there since a parton has freedom to decay at any point between $x_0$ and 
forward to our maximum distance $\frac{1}{Q}$. We have experimented with 
this factor and chosen some different 
distributions. One, perhaps more physical, is that instead of choosing a 
random number linearly to chose them from a exponential distribution. This 
agrees with a quantum mechanical picture, it is a high probability that it
 decays early but there exists a chance that it will live very long.

The second factor in Eq~(\ref{eq:evpos}) $\cos{\theta}$ is simply the
projection of the partons position to the $x$-axis as discussed in 
section~\ref{sec-begpos}. The angle  $\theta$ is randomly chosen.

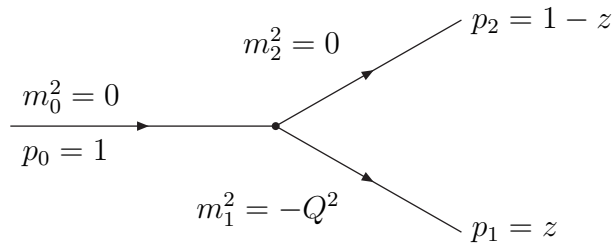
\begin{figure}[b]
\begin{center}   
\begin{picture}(400,100)(0,0)
\Vertex(200,50){1.5}
\ArrowLine(100,50)(200,50)
\ArrowLine(200,50)(270,90)
\ArrowLine(200,50)(270,10)
\Text(105,55)[lb]{$m^2_0=0$} 
\Text(105,45)[lt]{$p_0=1$}
\Text(275,90)[l]{$p_2=1-z$}
\Text(275,10)[l]{$p_1=z$}
\Text(225,75)[rb]{$m_2^2=0$}
\Text(225,25)[rt]{$m_1^2=-Q^2$}
\end{picture}
\end{center}  
\caption{The kinematics of a vertex}
\label{fig:kinpos}
\end{figure}

The third and last factor comes from the kinematics of the system, according to
Fig.~\ref{fig:kinpos}.
We have, from the conservation of energy and momentum, the following 
expressions:
\begin{eqnarray}
(E+p_z)_1=z(E+p_z)_0 
\label{cons1}  \\
(E+p_z)(E-p_z)=m^2+\pt^2
\label{cons2} 
\end{eqnarray}
Here $p_z$ is the direction of the incoming particle and $\pt$ is the 
direction orthogonal to this. If we apply Eq~(\ref{cons2}) on the initial state
(0) and the final state (1 and 2), we get
\begin{equation}
0 = \frac{-Q^2+\pt^2}{(E+p_z)_1} + \frac{\pt^2}{(E+p_z)_2} = 
\frac{-Q^2+\pt^2}{z(E+p_z)_0} + \frac{\pt^2}{(1-z)(E+p_z)_0}.
\end{equation}
In the last equality Eq~(\ref{cons1}) was used. From this expression it is
straight forward to get a relation between $\pt$ and $z$ according to:
\begin{equation}
\pt=\sqrt{1-z}\,\,Q.
\end{equation}
If a parton gets a large share, $z$, of the momentum it will not be
able to deviate much from the original partons direction. Therefore  
the distance it will travel according to Eq~(\ref{eq:evpos}) is small.
We have also studied results without this factor, and find that it is 
not critical for the qualitative results.

\subsection{The colour charge}
\label{subsec-colour}
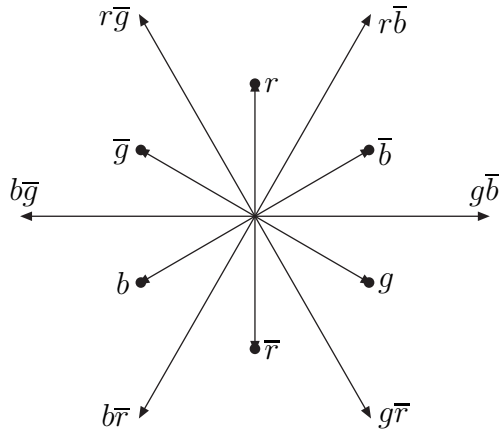
\begin{figure}[h]
\begin{center}   
\begin{picture}(200,200)(0,0) 
\Vertex(100,150){2}
\Vertex(100,50){2}
\Vertex(143,75){2}
\Vertex(143,125){2}
\Vertex(57,75){2}
\Vertex(57,125){2}
\LongArrow(100,100)(100,150)
\LongArrow(100,100)(100,50)
\LongArrow(100,100)(143,75)
\LongArrow(100,100)(143,125)
\LongArrow(100,100)(57,75)
\LongArrow(100,100)(57,125)
\LongArrow(100,100)(143,175)
\LongArrow(100,100)(143,25)
\LongArrow(100,100)(187,100)
\LongArrow(100,100)(13,100)
\LongArrow(100,100)(57,175)
\LongArrow(100,100)(57,25)
\Text(104,150)[l]{$r$}
\Text(104,50)[l]{$\overline{r}$} 
\Text(147,75)[l]{$g$} 
\Text(147,125)[l]{$\overline{b}$}    
\Text(53,75)[r]{$b$} 
\Text(53,125)[r]{$\overline{g}$}
\Text(147,175)[l]{$r\overline{b}$} 
\Text(147,25)[l]{$g\overline{r}$} 
\Text(187,104)[b]{$g\overline{b}$} 
\Text(13,104)[b]{$b\overline{g}$} 
\Text(53,175)[r]{$r\overline{g}$} 
\Text(53,25)[r]{$b\overline{r}$} 
\end{picture}
\end{center}  
\caption{The representation of colour charge in a 2-dimensional room}
\label{fig:colroom}
\end{figure}

Since the colour charge has 3 different charges (colours), instead of as in 
electromagnetism 2, you have to introduce one more dimension to represent
the colour space. That is, you will get a colour-plane, Fig~\ref{fig:colroom}, 
 where the different charges of the particles are added to each other 
as vectors. Though this is not completely true, since we have the same 
phenomena here as in the spin-space. Even though the $z$ component, $S_z$, of 
the spin is zero the total spin, $S$, can be a finite value, since $S_z$
is the projection of the spin vector on the $z$-axis. The same thing happens
in the colour space, we can have a colour neutral particle that is not a 
colour singlet. An example of this are the two colour neutral gluons
 (e.g.$\frac{1}{\sqrt{2}}(r\overline{r}-g\overline{g})$). We don't make
this distinction in our model though, and we treat the colour charge
as if it simply were vectors in a two dimensional space. So, e.g. you get the 
gluon $g\overline{b}$ charge by adding the vectors of the green charge and 
the antiblue charge, and so on.

This representation of the colour charge is not perfect, we have only 6 
gluons, as is seen in Fig.~\ref{fig:colroom}. The colour neutral gluons (2)
are not represented in this picture, since they would then coincide with
origo which would give a set of none-interacting gluons. Another thing that 
is not entirely correct is the length, or charge,
 that the gluons get. In this representation the length of the gluons is 
$\sqrt{3}$ while it should be $\sqrt{\frac{N_c}{C_F}}=
\sqrt{\frac{3}{4/3}}=\frac{3}{2}$.

For our simple model of the proton these small errors are not very important 
and this representation is adequate.\\

We generate the start configuration of the 5 partons as if we originally
had only 3 quarks that were colour neutral, e.g. one quark of each colour.
These emits 2 gluons randomly chosen (though not gluons that are colour 
neutral), according to $q\rightarrow qg$.

A possible start configuration could then be that the red quark emits a
gluon with $r\overline{b}$ and the blue emits a gluon  with $b\overline{g}$.
And we have 1 blue quark, 2 green quarks, 1 gluon with $r\overline{b}$ and 
another with $b\overline{g}$. If you add these together they will exactly 
cancel each other and we have a colour neutral configuration.

As we evolve our parton shower we follow the rules of QCD in each vertex.
Whenever there is freedom to choose colour is this done randomly. As 
long as we follow these rules we are guaranteed to have a colour neutral 
state. However, if we keep all partons, and evolve all this we will get
several hundreds of partons in our model and this is not desired. To avoid
this we have introduced a $x_{min}$, as mentioned before. Branchings which 
produces partons with $x<x_{min}$ are not allowed, but we keep all the partons
in the parton shower to keep the colour charge to be zero.

\subsection{Number of partons}
\label{subsec-numberofpartons}
Here we present a graph showing how the number of partons increase with
the virtuality, $Q$, for two different energy scales. We adjust the energy
by changing the variable $x_{min}$ and the lower  $x_{min}$ the higher
the energy.

%
\begin{figure}[h]
\begin{center}
\rotatebox{270}{\mbox{\epsfig{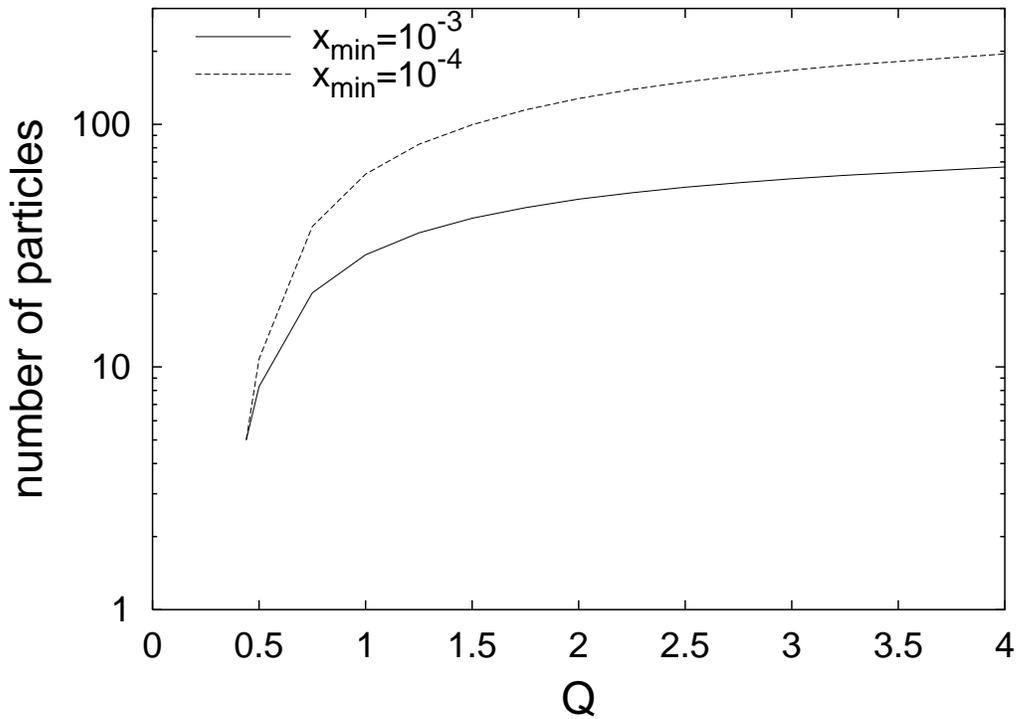}}}
\end{center}
\caption{How the number of partons increases with the virtuality, $Q$.}
\label{fig:antalp} 
\end{figure}
%

\subsection{A summary of the proton model}
\label{subsec-summod}
Our Model of the proton is made in an exclusive picture. We start at a low
$Q$ with only a few partons and let these partons evolve in an iterative
process to some $Q_{max}$. To reduce the number of produced partons we had to 
introduce a cutoff in the momentum space. The branchings that produces partons
 with a momentum under this cutoff were not allowed. Despite this rather
abrupt cut in the evolution of the parton shower we got a good distribution
of partons and their momenta. We have compared our results with GRV in
Fig.~\ref{fig:evmom1} to Fig.~\ref{fig:evmomlog} and the total parton 
shower coincides almost exactly.

We will now use this model to study some different phenomena, but mainly 
the screening effect.     

\clearpage
\section{A study of the Model}
\label{sec-study}
The primary goal of this section is to study how our model of the proton
interacts with a gluon, but also how the radius of the proton varies with
the energy of the system, section~\ref{sec-radius}. When we are sending a 
gluon against our proton
will we see that the screening effect comes into play and decreases the 
interaction. We are going to try some different start configurations and
see how large the screening effect is, and draw some conclusions of this,
section~\ref{subsec-screeningmod}.

\subsection{Introducing the Interference equation}
\label{subsec-interference}
As we discussed in sec~\ref{subsec-screening} is  the interesting 
equation when we are studying the interaction of an incoming gluon
against the proton the following:

\begin{equation}  
A=\frac{|\sum_{k=1}^n q_ke^{ipx_k}|^2}{\sum_{k=1}^n|q_k|^2}.
\label{eq:inteference}
\end{equation}

\begin{figure}[h]
\begin{center}
\rotatebox{270}{\mbox{\epsfig{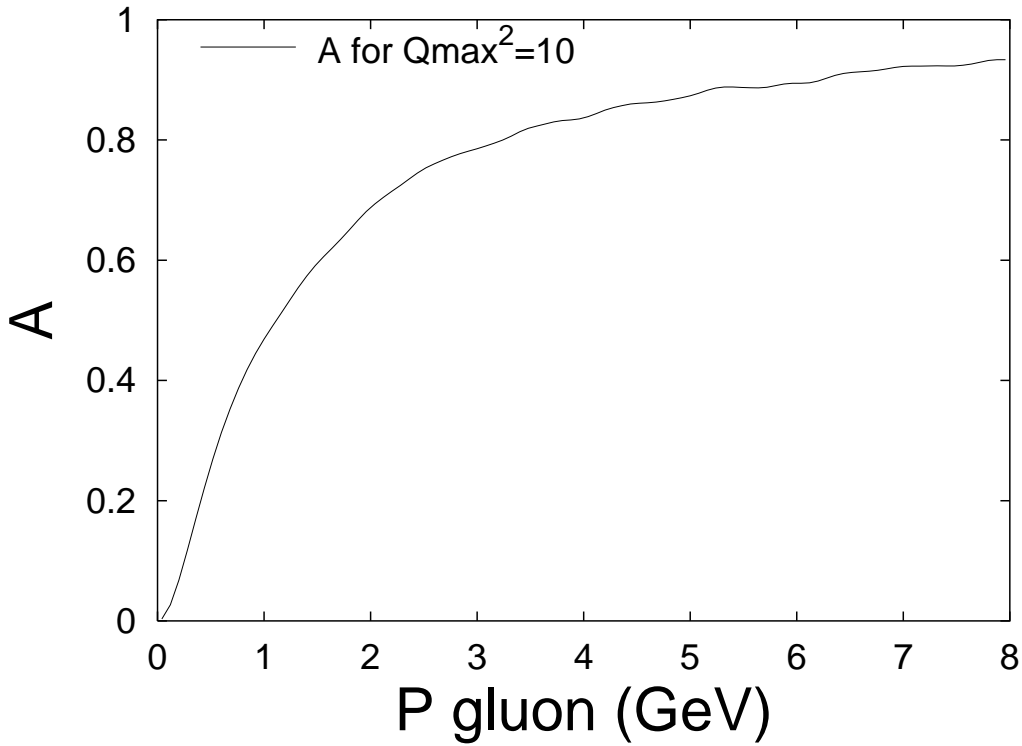}}}
\end{center}
\caption{The resolution of the proton increases with the gluon momentum and
therefore we approach an incoherent state. $x_{min}=10^{-3}$}
\label{fig:A}
\end{figure}

Since we in our model have assigned both position and colour charge to all
partons we can readily use this formula and study how $A$ varies when the
incoming gluon has different transverse momentum. According to the theoretical 
predictions we should have a curve that at low momentum for the gluon is
zero and then gradually increase to the maximum value 1 at very high 
momentum. This is exactly what we have, see Fig.~\ref{fig:A}.  
%
%

\clearpage
\subsection{A running Q}
\label{subsec-peqq}
In chapter~\ref{subsec-interference} we did the following: We first 
evolved the parton cascade according to the evolution equations. Then we
sent gluons with increasing momenta at the now evolved proton.

Another, more physical, approach would be to probe the proton with a gluon 
that had approximately the same value on $p$ as the partons has $Q$. As
we mentioned earlier (chapter~\ref{sec-theory}) , is $Q \sim \pt$. So
this procedure is built upon the principle that a probe can resolve best
at its own wavelength (or momentum). In practice this is done by 
continuously, as we evolve our parton cascade to higher $Q$, sending in a 
gluon that has the same $p$ as we have $Q$ at this point.    

%
\begin{figure}[h]
\begin{center}
\rotatebox{270}{\mbox{\epsfig{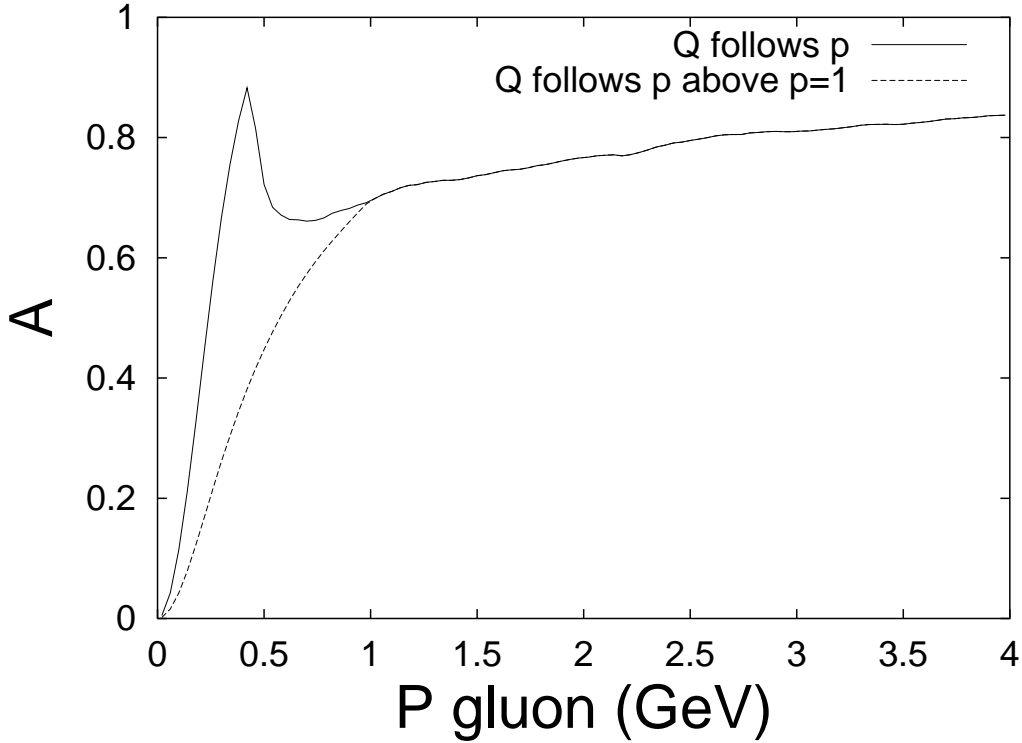}}}
\end{center}
\caption{The Interference function $A(p)$ when $Q$ follows the $p$ scale. In
the dashed curve is parton shower frozen at $Q=1$ and downward to avoid the
 peak at $p=0.44$. $x_{min}=10^{-3}$}
\label{fig:topp}
\end{figure}
%

The peak at $p=0.44$ in Fig.~\ref{fig:topp} on the upper curve should not be 
there. It arises from the fact that when we start our evolution at $Q=0.44$ we
have only 5 partons and these 5 partons have no tendency at all to cluster,
unlike the partons produced later in the evolution. Since the physics
 in this low region is little understood our start configurations can not be 
checked correctly and it is not certain that the ansatz with only few partons 
is applicable. What we have done is to take 
our distribution at a higher $Q$,
 in our case at $Q=1$, and use this distribution all the way down to $Q=0$. 
This is similar to what is often done to calculate jet cross-sections with
 parton distributions not defined below some $Q^2$. 
If we do this we get the dashed curve in Fig.~\ref{fig:topp} and the peak 
has disappeared.  
\clearpage 

\subsection{Introducing a variable $x_{min}$}
\label{subsec-xmin}
We introduced a cut in the momentum, $x_{min}$, and partons under this limit
was frozen. We did not discuss how this cut was chosen, nor its physical 
meaning. 

If we look at two incoming partons with a momentum of $x_1$ resp. $x_2$ 
we have:
\begin{equation}
\hat{s}=x_1 x_2 s,
\end{equation}
where $s$ is $E_{CM}^2$ and $\hat{s}$ is the energy of the subprocess 
squared. $\pt$ can at the most be half of the energy of the subprocess (since 
we have two partons) and this will mean that we get:
\begin{equation}
\pt^2 \leq \frac{\hat{s}}{4}=x_1 x_2 \frac{s}{4}\,\,\,\, 
\Longrightarrow \,\,\, \pt^2 \sim x_{min}^2 \frac{s}{4}, 
\end{equation}
that is, we have
\begin{equation}
x_{min}^2 \geq  \frac{4\pt^2}{s}.
\end{equation}
This is a somewhat crude calculation, but we see two things here. First, and
not so surprising, that the higher total energy we have the better 
we can resolve partons with small momentum, i.e we get a lower $x_{min}$. The
other thing is that, for a fixed $E_{CM}$, $x_{min}$ scales linearly with 
$\pt$. So, as we increase the virtuality scale, $Q$, we should continuously 
increase $x_{min}$.
%
\begin{figure}[h]
\begin{center}
\rotatebox{270}{\mbox{\epsfig{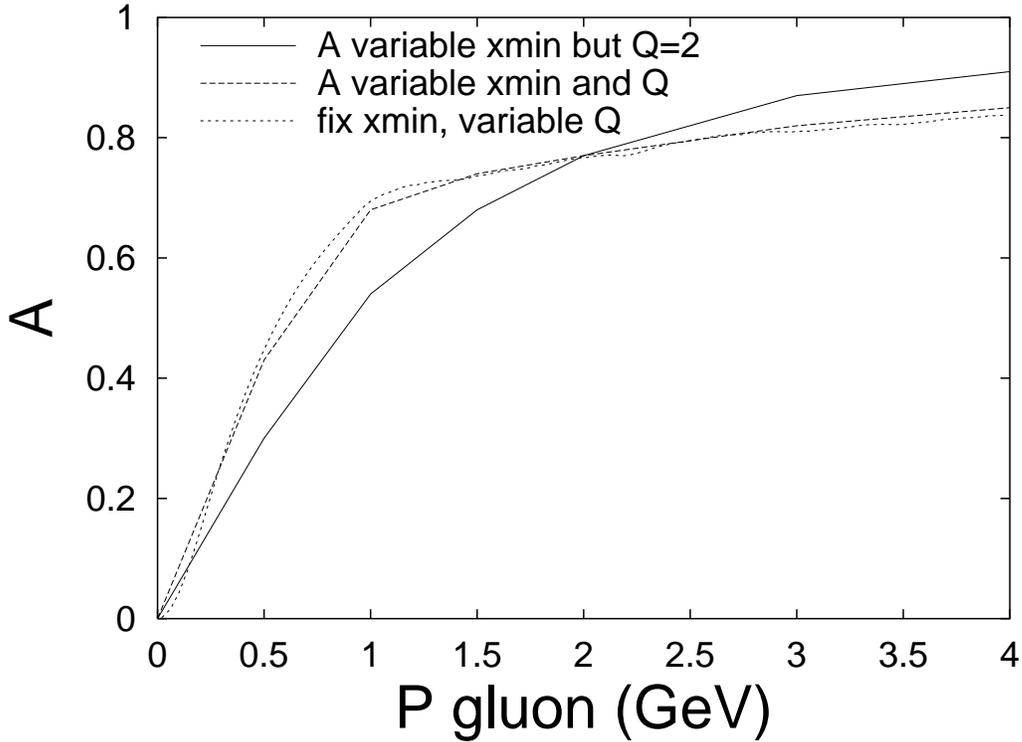}}}
\end{center}
\caption{Introducing a variable $x_{min}$. In the dashed curve is $Q$ running
with $p$ while in the other curve is $Q=2$. We have also included the curve 
for a fix $x_{min}=10^{-3}$.}
\label{fig:varxmin}
\end{figure}
%

In Fig.~\ref{fig:varxmin} we have that $x_{min}=10^{-3} \, \pt$. This factor 
$10^{-3}$ can be varied and the curves will be suppressed for smaller factors. 
In the dashed curve with a running $Q$ have we, as said in 
chapter~\ref{subsec-peqq}, frozen the parton distribution at $Q=1$ and 
downward. As can be seen in Fig.~\ref{fig:varxmin} this effect is small.

\subsection{The Screening Effect}
\label{subsec-screeningmod}
All the graphs in chapter~4 are a measure of the screening effect since
they all tell you how well the gluon can resolve the partons inside the 
proton. By parameterizing these curves by some function $S(\pt)$ you would
see the strength of the screening effect and how it increases with the energy.
As a reasonable guess, we put:
\begin{equation} 
S(\pt)=\frac{\pt^2}{\pt^2+p_{\bot 0}^2},
\label{S}
\end{equation}
where $p_{\bot 0}$ is a free parameter. As long as $\pt$ is large is $S$
equal to 1, it is only when $\pt$ approaches zero that our correction term
comes into play.

\begin{figure}[h]
\begin{center}
\rotatebox{270}{\mbox{\epsfig{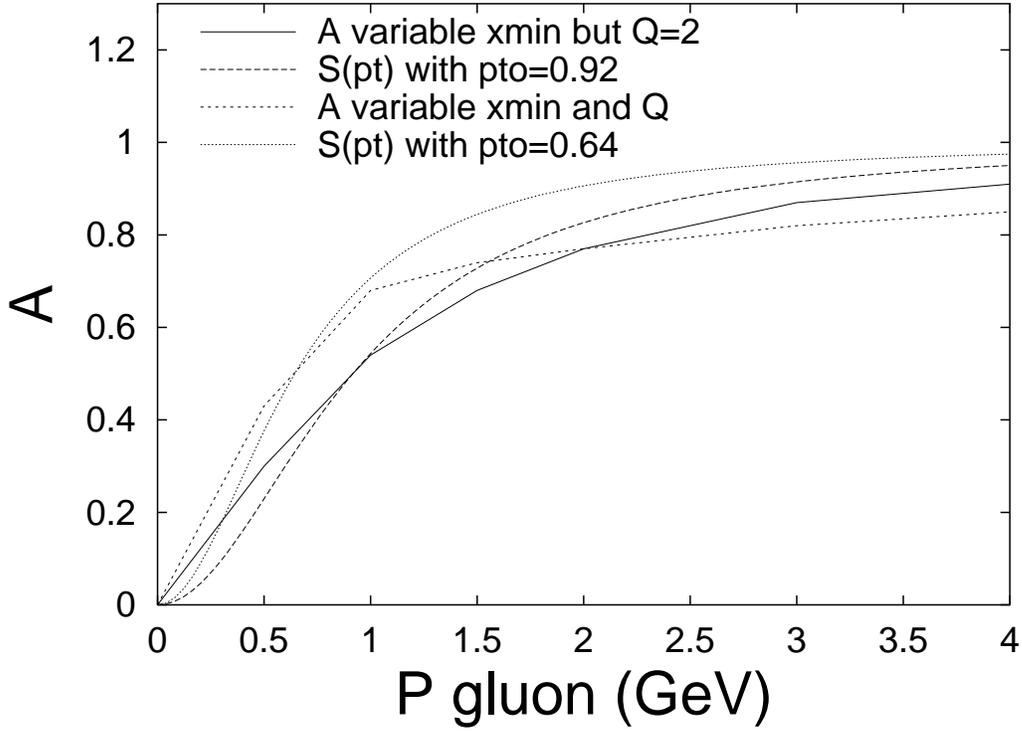}}}
\end{center}
\caption{The screening function S($\pt$) fitted to A. Our parameterization of 
the screening function can best be fitted to the curve with a fix Q.}
\label{fig:anpassning} 
\end{figure}

In the expression for the cross-section $\sigma_{jet}$ we had
\begin{equation}
\mrm{d}\sigma_{jet} = \int \mrm{d}x_{1}f(x_{1},\pt^2)\int \mrm{d}x_{2}
f(x_{2},\pt^2)
\int \frac{\mrm{d}\hat{\sigma}}{\mrm{d}\pt^2}\mrm{d}\pt^2,
\label{cross-sectj1}
\end{equation}
or, if we rewrite it a little
\begin{equation}
\frac{\mrm{d}\sigma_{jet}}{\mrm{d}\pt^2} =\frac{\mrm{d}\hat{\sigma}}
{\mrm{d}\pt^2} \int \mrm{d}x_{1}f(x_{1},\pt^2)
\int \mrm{d}x_{2}f(x_{2},\pt^2).
\label{cross-sectj2}
\end{equation}
It is the first term in this expression that has a singularity as 
$\pt\rightarrow0$ since it goes like $1/\pt^4$. By introducing $S(\pt)$ 
(squared since we have two partons) into
Eq~(\ref{cross-sectj2}) this singularity disappears and we get a finite 
value:
\begin{eqnarray}
\frac{\mrm{d}\sigma_{jet}}{\mrm{d}\pt^2} & = & S(\pt)^2 
\frac{\mrm{d}\hat{\sigma}}{\mrm{d}\pt^2} \int \mrm{d}x_{1}f(x_{1},\pt^2)
\int \mrm{d}x_{2}f(x_{2},\pt^2)   \nonumber \\
& = & \frac{\pt^4}{(\pt^2+p_{\bot 0}^2)^2}
\frac{\mrm{d}\hat{\sigma}}{\mrm{d}\pt^2} \int \mrm{d}x_{1}f(x_{1},\pt^2)
\int \mrm{d}x_{2}f(x_{2},\pt^2)   \nonumber \\
& \sim & \frac{1}{(\pt^2+p_{\bot 0}^2)^2}\int \mrm{d}x_{1}f(x_{1},\pt^2)
\int \mrm{d}x_{2}f(x_{2},\pt^2)   \\
\label{cross-sectj3} \nonumber
\end{eqnarray}
When introducing the screening function $S(p_{\bot})$ we adjust the parameter
$p_{\bot 0}$ e.g., so that $S(p_{\bot})$ coincides 
with our curve at $A=0.5$, see
Fig.~\ref{fig:anpassning}. By doing this for different energies we can plot
how the screening effect varies by the energy. We change the energy of the
parton shower by changing $x_{min}$ since we had the relation that 
$x_{min} \sim 1/E_{CM}$. 

The curve that has a fix $Q$ can better be fitted with our screening function
than the one with a running $Q$, see Fig.~\ref{fig:anpassning}. This indicates
a shortcoming either of the model or in the ansatz of $S$. We will use
a fix $Q$ mostly, to allow a more sensible comparison.

\begin{figure}[h]
\begin{center}
\rotatebox{270}{\mbox{\epsfig{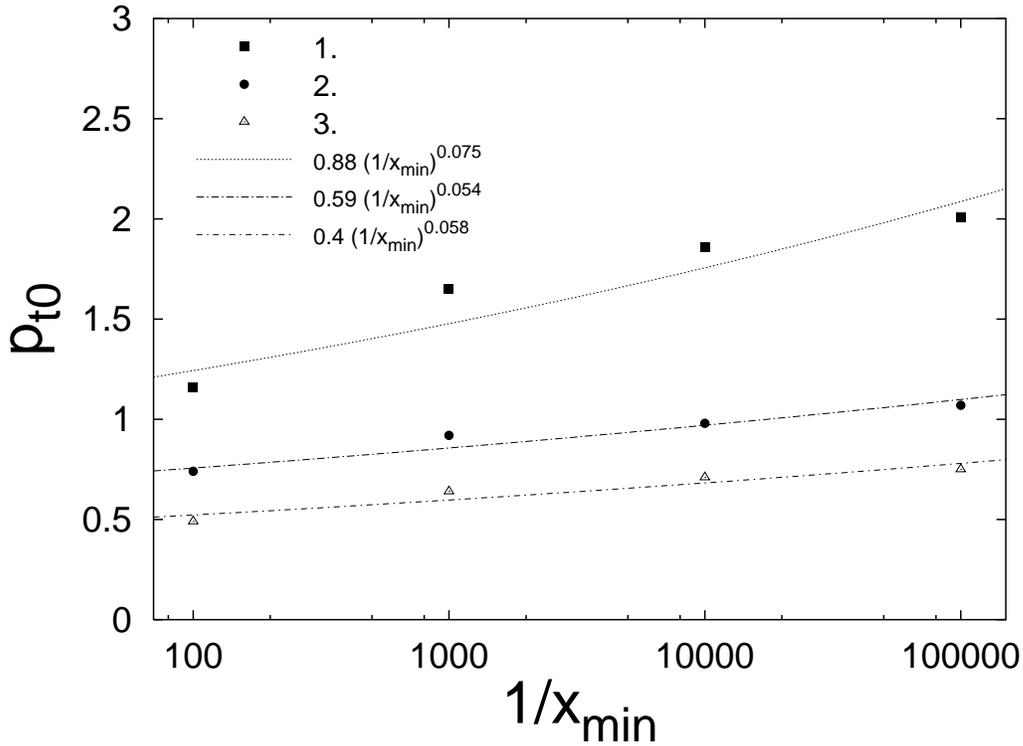}}}
\end{center}
\caption{How $p_{\bot 0}$ varies with the energy for different start 
configurations. Scenarios 1,2 and 3 according to the notation in the text.}
\label{fig:pto_vs_xmin} 
\end{figure}

By plotting $p_{\bot 0}$ versus $1/x_{min}$ we get a good measure of
how the screening effect varies by the energy. We have, in 
Fig~\ref{fig:pto_vs_xmin} done this for some different start 
configurations and also some different evolution
conditions. Since we do not know how much the partons separate themselves in a
branching we will also look at events where the splitting is twice the size
of  Eq~(\ref{eq:evpos}) 
\begin{enumerate}
   \item A fix $Q=2$ and a branching  according to
         Eq~(\ref{eq:evpos})   
   \item A fix $Q=2$ and a  branching twice the size of
         Eq~(\ref{eq:evpos})
   \item A running $Q$ and a  branching twice the size of
         Eq~(\ref{eq:evpos}) 
\end{enumerate}
We see in Fig.~\ref{fig:pto_vs_xmin} that $p_{\bot 0}$ increases  
as the energy of the system is increased. By this we get that $\sigma_{jet}$
increase slower than for a fixed  $p_{\bot 0}$ and we get a nicer behavior 
than in Fig~\ref{fig:traffyta}.

%
%
%
%
\clearpage
\subsection{The Radius of the Proton in our Model}
\label{sec-radius}
As another application of our model we are going to look at how the radius
of our model of the proton evolves as we increase the energy of the system.

We do this by plotting the parton density function $f(r)$:
\begin{equation}
f(r)=\frac{1}{r}\frac{\mrm{d}n_{partons}}{\mrm{d}r}=\frac{\mrm{d}n_{partons}}
{\mrm{d}^2r},
\end{equation} 
the phase-space factor $1/r$ is introduced to avoid having a vanishing 
distribution at $r=0$.
The center, from which the distance $r$ is calculated, is simply the total
average of the position of all partons.

%
\begin{figure}[h]
\begin{center}
\rotatebox{270}{\mbox{\epsfig{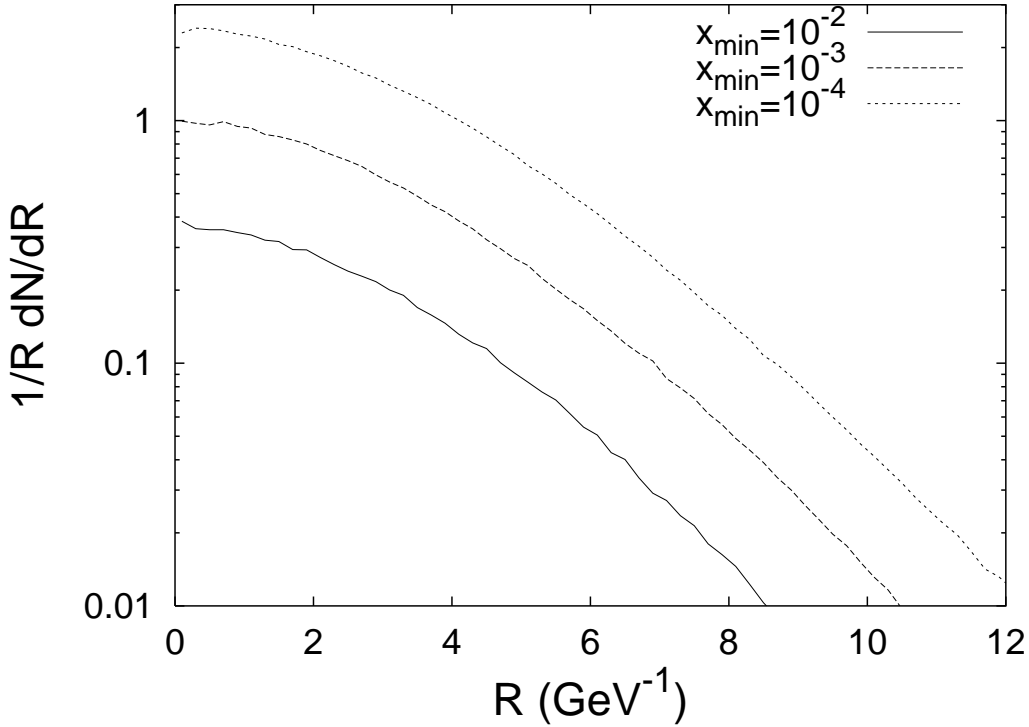}}}
\end{center}
\caption{How the radius of our proton model varies with the energy.}
\label{fig:medelradie} 
\end{figure}
%

By defining some lowest density of the proton it is easy to define a radius
as the distance from origo to the point on the curve at this lowest 
density.
To vary the energy we once again changes the parameter $x_{min}$, see 
Fig~\ref{fig:medelradie}. Here the density function is plotted for some 
different $x_{min}$. As is expected, the radius increases with the energy.  
We see that the curves in Fig~\ref{fig:medelradie} are almost parallel to
each other. This means that the partons stays rather close to their original
parton mother. Though a closer examination of Fig~\ref{fig:medelradie} shows
that the curves spreads towards larger $R$. This is an indication of
some diffusion, i.e. partons that have traveled far from their
original position.  
\clearpage
\section{Summary and Conclusions}
\label{sec-sum}
This paper is divided in two main parts:
\begin{enumerate}
   \item The Proton Model
   \item The Studies of the Model
\end{enumerate}
In the first part we describe how the model is constructed and which 
conditions we have made upon it.

The model is iterative in its structure; we start with a configuration of only
5 partons. These are then evolved towards higher virtuality scales $Q$  to 
some $Q_{max}$ where we stop the evolution. The evolution is done in an
exclusive picture i.e. we follow each parton by simulating the probability 
for a branching by random numbers. To know the probability for a branching we 
use the Altarelli-Parisi splitting kernels. 

If we allowed all partons there would be infinitely many in our model, and
this is  not possible. To avoid this we introduced a cutoff in momentum space.
The branchings that produces partons with a momentum below this $x_{min}$
were not allowed, all the partons are
still kept into the parton shower though. Despite this rather abrupt cutoff,
the distribution of the partons is not changed significantly.

We have also made the model so that it is always colour neutral. This is done 
by having a colour neutral start configuration. Then, as we evolve, the rules 
of QCD are followed and no partons are thrown away.

All partons were assigned a transverse position in the coordinate space  and a
longitudinal value of the momentum. This is done to try to avoid difficulties
from the Heisenberg uncertainty principle.

We have compared our model to the model by the GRV group, 
see Fig.~\ref{fig:evmom1} to Fig.~\ref{fig:evmomlog}. Our model cannot compete
with GRV but we have a rather good agreement between the two models, especially
at high $Q$.\\    

In the second part of this paper we studied the so called screening effect
with our model. The screening effect is that the partons will screen each
other so the charge seen from outside the proton will be lower than the 
naive incoherent sum. By using
our model as a target we sent gluons with different momentum at the proton.

We made some different assumptions of how the gluon interact with the proton
model:

\begin{itemize}
  \item A running $Q$: \, That the incoming gluon will react mostly with 
        partons with $Q$ (and thereby $p_{\bot}$) equal to the gluon momentum
  \item A running $x_{min}$: \, The cutoff variable  $x_{min}$ should vary
        with the $Q$ scale.
  \item The distance between the partons: \, When a branching occurs  the 
        distance between the daughters are decided by Eq~(\ref{eq:evpos}), this
        is varied 
\end{itemize}

The different curves of this is presented in Fig.~\ref{fig:pto_vs_xmin}. We
see that there are some differences, but the qualitative behavior is the same.
 We clearly have the screening effect and the $\sigma_{jet}$ will be better
behaved than in Fig.~\ref{fig:traffyta}.

\subsection*{Acknowledgement}
I would first of all like to thank my supervisor Torbj\"orn Sj\"ostrand for 
his support and patience as well as for introducing me to this subject. I would
also like to thank the rest of the department who have helped me during this
time. And of course my family and friends for their support and encouragement.

\pagebreak

\end{document}